\documentclass[12pt]{article}
\usepackage{graphicx}
\usepackage{amsfonts}
\usepackage{amssymb}
\usepackage{float}
\usepackage[footnotesize]{caption2}
\usepackage{mathrsfs}
\usepackage{amsmath}
\usepackage{indentfirst}
\usepackage{hyperref}
\usepackage{cite}
\usepackage{authblk}

\begin{document}
\renewcommand{\thefootnote}{\fnsymbol{footnote}}
\title{Hawking tunneling radiation with thermodynamic pressure}
\author[1]{\small Cheng Hu }
\author[2]{\small Xiao-Xiong Zeng \thanks{corremail address: xxzengphysics@163.com} }

\affil[1]{\small College of Basic Teaching, Chengdu Neusoft University, Chengdu 610000, China}
\affil[2]{\small College of Physics and Electronic Engineering, Chongqing Normal University, Chongqing 401331, China}

\date{}
\maketitle
\begin{abstract}

\setlength{\parindent}{0pt} \setlength{\parskip}{1.5ex plus 0.5ex minus 0.2ex} %\noindent
Hawking radiation elucidates black holes as quantum thermodynamic systems, thereby establishing a conceptual bridge between general relativity and quantum mechanics through particle emission phenomena. While conventional theoretical frameworks predominantly focus on classical spacetime configurations, recent advancements in Extended Phase Space thermodynamics have redefined cosmological parameters (such as the $\Lambda$-term) as dynamic variables. Notably, the thermodynamics of Anti-de Sitter (AdS) black holes has been successfully extended to incorporate thermodynamic pressure $P$. Within this extended phase space framework, although numerous intriguing physical phenomena have been identified, the tunneling mechanism of particles incorporating pressure and volume remains unexplored.
This study investigates Hawking radiation through particle tunneling in Schwarzschild Anti-de Sitter black holes within the extended phase space, where the thermodynamic pressure $P$ is introduced via a dynamic cosmological constant $\Lambda$. By employing semi-classical tunneling calculations with self-gravitation corrections, we demonstrate that emission probabilities exhibit a direct correlation with variations in Bekenstein-Hawking entropy. Significantly, the radiation spectrum deviates from pure thermality, aligning with unitary quantum evolution while maintaining consistency with standard phase space results. Moreover, through thermodynamic analysis, we have verified that the emission rate of particles is related to the difference in Bekenstein-Hawking entropy of the emitted particles before and after they tunnel through the potential barrier. These findings establish particle tunneling as a unified probe of quantum gravitational effects in black hole thermodynamics.


\end{abstract}

\thispagestyle{empty}

%=========+=========+=========+=========+=========+=========+=========+=========
\newpage

\section{Introduction}

In 2019, the Event Horizon Telescope (EHT) collaboration achieved a historic milestone by capturing and publicly releasing the first direct visual evidence of the supermassive black hole at the center of the M87 galaxy, located 55 million light years from Earth \cite{bib1}. In fact, as far back as the 1970s, Stephen Hawking made a pioneering discovery: the size of a black hole's event horizon does not decrease\cite{bib2,bib3}. Ever since then, it has been widely accepted that a black hole can be regarded as an ordinary thermodynamic system. In this thermodynamic system, a black hole possesses an event horizon beyond which no particle, not even light, can escape the intense gravitational pull. This characteristic implies that no energy or matter within the black hole can reach an external observer. Nevertheless, when quantum effects are taken into account, a small amount of energy can be radiated beyond the black hole's spacetime boundary. Previous research has demonstrated that the temperature of a black hole can be defined based on the radiant energy it emits. At the same time, black holes can also be regarded as thermodynamic systems with Hawking temperatures\cite{bib4}. Furthermore, black holes have irreducible mass, a property that increases in irreversible processes. The irreducible mass is similar to entropy in ordinary thermodynamic systems. This is the Bekenstein-Hawking entropy of a black hole\cite{bib5,bib6}, which is proportional to the area of horizon. The thermodynamic law of black hole as a thermal system is established by using temperature and entropy. In normal phase space, the conserved quantities of the black hole thermodynamic system, such as charge, angular momentum, and thermodynamic potential of the system, are also defined on the event horizon. In other words, based on the black hole horizon, the thermodynamics of black holes is established.
Afterwards, Hawking radiation on the thermal properties of black holes in particular has been extensively studied, such as, the Thermal Green's function method\cite{bib7}, the Euclidean action integral approach\cite{bib8}, the Second quantization method\cite{bib9}, the Punsly's approach\cite{bib10}, Damour-Ruffini Method\cite{bib11}, and so on\cite{bib12,bib13,bib14,bib15,bib16,bib17,bib18,bib19,bib20}. However, in these numerous studies, the radiation spectra obtained are purely thermal because the self-gravitational interactions of the thermodynamic system of the black holes are not taken into account.

In 2000, F. Wilczek and M. Parick \cite{bib21,bib22} introduced an innovative framework to investigate Hawking radiation through a quantum tunneling formulation. Within this theoretical framework, the black hole event horizon is conceptualized as an effective potential barrier in quantum mechanical tunneling processes. By considering the energy conservation and the spacetime background structure, the position of the potential barriers is given, and the tunneling probability of the black hole radiation is obtained. This analytical approach yielded a modified thermal emission spectrum and prove that Hawking radiation of black hole is not a pure thermal spectrum. Due to the generalizability of this approach, subsequent research rapidly expanded its application to various classes of black hole solutions \cite{bib23,bib24,bib25,bib26,bib27,bib28,bib29,bib30,bib31,bib32,bib33,bib34,bib35,bib36,bib37,bib38,bib39,bib40,bib41,bib42,bib43,bib44,bib45,bib46,bib47,bib48,bib49,bib50,bib51,bib52,bib53}. 
Notably, when implementing this methodology in de Sitter and Anti-de Sitter (Ads) black hole spacetimes, prior investigations predominantly treated the cosmological parameter $\Lambda$ as a static quantity.\cite{bib44,bib45,bib46,bib47,bib48,bib49,bib50,bib51,bib52,bib53}.
Recently, however, researchers have noticed that if cosmological parameters are considered as dynamic parameters of the thermodynamic system, it will be interesting to explore their contribution and influence on the thermodynamic system of black holes. If the cosmological constant is considered as dynamic parameters of the thermodynamic system of black holes, in the extended phase space, whether the law of black hole thermodynamics is valid or not, and whether the cosmological supervision assumption is valid or not are worth exploring and investigating in depth, to describe the thermodynamic properties of black holes in a more comprehensive way\cite{bib54,bib55,bib56,bib57,bib58}.

Caldarelli\cite{bib59} pioneered the elevation of the cosmological constant to a variable state parameter of the thermodynamic system of black holes and systematically investigated the thermodynamic properties of Kerr-Newman-AdS black holes. Hendi\cite{bib60} directly called the phase space containing the cosmological constant($M, Q, \Lambda$ ) as an extended phase space and further studied the thermodynamic phase transition of a class of charged black holes. In addition, some scholars have proposed a volume of a black hole which can be defined as a conjugate variable of the thermodynamic system, claiming that the role of the cosmological constant turns out to be a natural candidate for pressure.
By considering $\Lambda$ as a dynamic thermodynamic variable, the new thermodynamic quantity, namely the thermodynamic pressure and volume of the black hole, has been introduced into the laws of black hole thermodynamics\cite{bib61,bib62}. Here, the thermodynamic press is $P=-\Lambda/8\pi$ and the volume is its conjugate quantity is $ V=(\partial M/\partial P)S,Q,J$, and the black hole mass $M$ is interpreted as thermodynamic enthalpy $\mathcal{H}$ rather than internal energy $U$ \cite{bib55}, which is defined as entropy versus the pressure variable.
In fact, from the general relativistic point of view, a negative cosmological parameter is sometimes puzzling when interpreted as a conjugate dynamic parameter of a thermodynamic volume, i.e., the cosmological constant is regarded as a dynamic parameter of a black hole thermodynamic system. But more than that, the theory has aroused great interest and widespread concern. Based on this theory, many interesting phenomena, such as van der Waals phase transition of black hole thermodynamic system, weak cosmic supervision hypothesis, repulsive interaction of black hole microstructure and whether the Smarr relation is consistent with the first law of thermodynamics have been studied extensively, and many meaningful results have been obtained\cite{bib63,bib64,bib65,bib66,bib67,bib68}. These studies show that the introduction of pressure and volume terms into the normal phase space enriches the phase structure space of the black hole thermodynamic system. In this extended phase space, the thermodynamic laws of black holes are still valid, and the weak cosmic supervision hypothesis is also valid, which has important practical significance. 
In the extended phase space, the first law of black hole thermodynamics is modified as $dM=TdS+VdP+\Phi dQ+\Omega dJ$\cite{bib68}. 

This correction complicates the thermodynamic properties of black holes and provides a new perspective on the study of Hawking radiation. The mechanism of Hawking radiation is similar to the conventional case, but the radiation spectrum and the evaporation process of the black hole may be affected due to the introduction of variables such as pressure and volume. Specifically, the Hawking radiation in the extended phase space depends not only on the mass of the black hole, may also be related to the pressure and volume, and this dependence makes the intensity and time evolution process of Hawking radiation different from the classical case, and the results are unknown to be tested.
When we investigate the tunneling process of a particle in the extended phase space, because the cosmological parameter $\Lambda$ is taken into account, the thermodynamic pressure and volume should not be ignored.
In fact, when a black hole radiates particles, its mass must decrease, the black hole is compressed, and the thermodynamic pressure and volume change. 
Within the extended phase space, the mass $M$ of an AdS black hole should be reinterpreted as its enthalpy $\mathcal{H}$ , which satisfies the thermodynamic relation $\mathcal{H}=M= U+P V$ (where $U$ denotes the internal energy) \cite{bib55}.
In view of this, by considering the self-gravitation, our motivation is straightforward in this paper, i.e. to use the Null Geodesic method to carefully investigate the Hawking radiation as tunneling of the Schwarzschild Anti-de Sitter black hole in the extended phase space.

The remainder of this paper is organized as follows.
In Sect.\ref{sec:2}, we discuss the extended phase space thermodynamics of the AdS Sitter black hole, where the cosmological constants are viewed as dynamical variable quantities.
Subsequently, we obtain the geodesic equation of tunneling particles by using Lagrangian analysis of the action.
In sect.\ref{sec:3},  we investigate the Hawking radiation as tunneling of the black hole with the thermodynamics variables. We discuss the tunneling process through thermodynamic analysis in sect.\ref{sec:4}.
Finally, we summarize and discuss our results in Sect.\ref{sec:5}.

%=========+=========+=========+=========+=========+=========+=========+=========
\section{Thermodynamics and Geodesic equation with pressure}\label{sec:2}
In this section, we focus on asymptotically AdS black holes within the framework of a negative cosmological constant $\Lambda$. While a concise overview is provided here, further discussions on the thermodynamics of AdS black holes can be found in Ref\cite{bib59}. For such black holes, it is customary to define the thermodynamic pressure as $P = -\Lambda/8\pi$, and it is clear that asymptotically flat black holes are the special case where the pressure is constant to zero, and once the pressure is introduced into black hole thermodynamics, we wish to consider thermodynamics that is not banal, as can be seen by the definition of the pressure, P, as a variable quantity. That is, the cosmological constant $\Lambda$ should be a variable physical quantity in the broad sense \cite{bib58}. The cosmological constant as a variable quantity is physically very interesting, and not only that, it is reasonable and even necessary that the cosmological constant undergoes change.
\subsection{thermodynamics with pressure}
For simplicity, we employ a spherically symmetric AdS black hole to discuss the tunneling behavior across the event horizon. In general, a spherically symmetric black hole can be expressed as
\begin{equation}
d s^2={-f(r)}{d t_s}^2+{f(r)^{-1}}{d r^2}+ {r^2}(d\theta^2+\sin^2\theta{d\phi^2}),
\end{equation}
where the metric function $f(r)$ is
\begin{equation}
f(r) = 1-\frac{2 M}{r}-\frac{\Lambda}{3} r^{2}, 
\end{equation}
here, the parameter $M$ denotes the black hole mass and $ \Lambda $ is the cosmological constant.
The cosmological constant $\Lambda$ was interpreted as the thermodynamic pressure in the extended phase space\cite{bib62}
\begin{equation}
P=-\frac{1}{8\pi}\Lambda=\frac{3}{8\pi l^2}.
\end{equation}
And, the metric function $f(r)$ can be modified as
\begin{equation}\label{q4}
f(r) = 1-\frac{2 M}{r}+\frac{8 \pi P}{3} r^{2}.
\end{equation}
The Black hole mass is given by solving $f(r) = 0$ in the equation, which are given below:
\begin{equation}\label{q6}
M = \frac{1}{2}(r_{H}+\frac{8}{3} \pi P r_{H}^{3}),
\end{equation}
where $r_H$ is the radius of the event horizon of the black hole, the black hole that is taken to be the largest real positive root of $f (r) = 0$. From this, the thermodynamic volume can be derived as
\begin{equation}
 V=(\frac{\partial M}{\partial P})=\frac{4\pi r_h^3}{3},
\end{equation}
Accordingly, the phase space of AdS black holes is extended. In the extended phase space, the first law of thermodynamics should be written\cite{bib55}
\begin{equation}
    \begin{array}{ccc}dM&=&TdS+VdP.\end{array}
\end{equation}
At the event horizon, one can also derive the volume as well as the entropy of the black hole, namely\cite{bib5,bib6,bib56}
\begin{equation}
S = \frac{A}{4} = \pi r_{H}^{2}.
\end{equation}
%=========+=========+=========+=========+=========+=========+=========+=========
\subsection{Geodesic equation of particles}
The event horizon of the black hole is located at $r = r_+$, where $r_+$ represents the largest root of $f(r) = 0$, and at this radius, the standard coordinates become singular. To properly describe particle trajectories across the event horizon, one must eliminate these coordinate singularities. This can be achieved by introducing the Painlev\'{e} coordinate transformation\cite{bib69}.
\begin{equation}
    dt_s=dt+\frac{\sqrt{1-f(r)}}{f(r)}dr,
\end{equation}
where ${t_s}$ denotes the Schwarzschild time and ${t}$ denotes the Painev\'{e} time. In this case, the Painev\'{e}-like line element of the Anti-Sitter black hole reads
\begin{equation}
    ds^2=-f(r)dt^2+2\sqrt{1-f(r)}drdt+dr^2+r^2(d\theta^2+\sin^2\theta d\phi^2).
\end{equation}
To describe the geodesic equation, we start with the action $I$, which governs the motion of a particle or photon in a curved spacetime. The action is given by:
\begin{equation}
    I=\int\mathcal{L}d\lambda,
\end{equation}
where $\mathcal{L}$ is the Lagrangian quantity, should be written as
\begin{equation}
    \begin{aligned}\mathcal{L}&=\frac{1}{2}g_{\mu\nu}\dot{x^\mu}\dot{x^\nu}\\&=\frac{1}{2}\left[-f(r)\dot{t}^2-2\sqrt{1-f(r)}\dot{t}\dot{r}+\dot{r}^2+r^2\dot{\theta}^2+r^2\sin^2\theta\dot{\phi}^2\right],\end{aligned}
\end{equation}
here $\dot{x^{\mu}}$ represents the derivative with respect to the affine parameter $t $. Based on the definition of the generalized momenta, namely$\mathcal{P}_\alpha=\frac{\partial\mathcal{L}}{\partial\dot{x_\alpha}}$ , we have
\begin{equation}\label{q12}
\begin{aligned}
  \mathcal{P}_{t} & =\frac{\partial\mathcal{L}}{\dot{t}}=-f(r)\dot{t}-\sqrt{1-f(r)}\dot{r}, & \mathcal{P}_r & =\frac{\partial\mathcal{L}}{\dot{r}}=-\sqrt{1-f(r)}\dot{t}+\dot{r} \\
  \mathcal{P}_{\theta} & =\frac{\partial\mathcal{L}}{\dot{\theta}}=r^2\dot{\theta}, & \mathcal{P}_\phi & =\frac{\partial\mathcal{L}}{\dot{\phi}}=r^2\sin^2\theta\dot{\phi}
\end{aligned}.\end{equation}
Consider the 4-velocity normalization condition, it is easy to see that $t$ and $\phi$ are the cyclic coordinates, so we have $\frac{d\mathcal{P}_t}{d\tau}=\frac{\partial\mathcal{L}}{\partial t}=0,\frac{d\mathcal{P}_\phi}{d\tau}=\frac{\partial\mathcal{L}}{\partial\phi}=0$. This means,
\begin{equation}\label{q13}
 \mathcal{P}_t=-f(r)\dot{t}-\sqrt{1-f(r)\dot{r}}=E=Constant,\quad\mathcal{P}_\phi=r^2\sin^2\theta\dot{\phi}=L=Constant,
\end{equation} 
here, we consider the geodesic in an invariant plane $\theta_1=\pi/2$ without loss of generality. Then, the value of $\dot{t}$ and $\dot{r}$ can be obtained as the following form  Eq.(\ref{q14})by using Eqs.(\ref{q12}) and (\ref{q13}), it is
\begin{equation}\label{q14}
\begin{aligned}
&\dot{t}=\frac{dt}{d\tau}=\pm\sqrt{E^2-f(r)\left(k+\frac{L^2}{r^2}\right)}\\
&\dot{r}=\frac{dr}{d\tau}=\frac{1}{f(r)}\left[-E\pm\sqrt{1-f(r)}\sqrt{E^2-f(r)\left(k+\frac{L^2}{r^2}\right)}\right]
\end{aligned}.
\end{equation}
Using Eq.(\ref{q14}), it is easy to calculate the geodesic equation, which is
\begin{equation}\label{q15}
\bar{r}=\frac{dr}{dt}=\frac{\dot{r}}{\dot{t}}=\frac{f(r)}{-\sqrt{1-f(r)}\pm\frac{E}{\sqrt{E^2-f(r)(k+L^2)/r^2}}},
\end{equation}
where the sign $\pm$ is related to the outgoing(ingoing) radial motion of the particle. When one considered  the facts that the s-wave approximation $(L = 0)$ and $k = 0$, Eq.(\ref{q15}) will reduce to
\begin{equation}\label{q17}
    \bar{r}=\pm1-\sqrt{1-f(r)}.
\end{equation}
In view of this, the geodesic equation (\ref{q17}), which applies to massless particles, has been derived by Lagrangian analysis. In the next section, we will use this equation to systematically study the tunneling radiation from the Schwarzschild anti-de Sitter black hole in an extended phase-space framework.

%=========+=========+=========+=========+=========+=========+=========+=========
\section{Tunneling with thermodynamic pressure}\label{sec:3}
In this section, we will study Hawking radiation as a semi-classical tunneling process considering the thermodynamic pressure and volume. In the vicinity of the black hole event horizon (the boundary is very close to the interior), positive and negative pairs of particles are generated because of the vacuum rise and fall of the quantum effect. The negative energy particles are absorbed by the black hole, and the positive energy particles are radiated to infinity through the field of view, thus reducing the black hole energy and pressure.
When the self-gravitational effect of the emission particles is considered, Eqs. (\ref{q4}) and (\ref{q17}) will be modified accordingly. Considering the emission particles as $S-$waves, both the interior and exterior of the black hole are anti-de Sitter spacetime, but the mass parameters of the interior and exterior are different when fixing the mass of the spacetime as a whole and allowing the mass of the black hole to vary. 
In the previous discussion \cite{bib17}, we have extended the achievements in phase space to an extended phase space involving thermodynamic pressure and volume. Different from the simplified treatment in previous literatures that equate mass with internal energy, in the framework of the extended phase space where thermodynamic pressure (related to the cosmological constant $\Lambda$) and volume are introduced, the internal energy $U$ of the black hole can no longer be simply interpreted as the black hole mass $M$. This is because here the mass needs to be redefined as the thermodynamic enthalpy $\mathcal{H}$ that includes the pressure-volume contribution, which satisfies $\mathcal{H} = M = U + PV$. This correction implies that the energy composition of the black hole not only includes the internal energy corresponding to the internal degrees of freedom but also must incorporate the energy term from the interaction between the spacetime background pressure and the black hole volume.
In the case of energy conservation, the mass $M$ and pressure $P$ of the black hole will be transformed accordingly: $M\longrightarrow M-\omega^{\prime}$ and $P\longrightarrow P-p^{\prime}$. 
Therefore, the Bekenstein-Hawking entropy difference of the black hole obtained in the extended phase space should be rewritten as $\Delta S = S(M-\omega^{\prime}, P-p^{\prime}) - S(M,P)$ rather than $\Delta S = S(M-\omega) - S(M)$. The expression of $f(r)$ for the black hole can be modified as
\begin{equation}
    f^{\prime}(r)=1-\frac{2(M-\omega^{\prime})}{r}+\frac{8\pi(P-p^{\prime})}{3}r^2.
\end{equation}
Meanwhile the geodesics equation of the particles becomes as
\begin{equation}\label{q19}
    \bar{r}=1-\sqrt{\frac{2(M-\omega^{\prime})}{r}-\frac{8\pi(P-p^{\prime})}{3}r^2}.
\end{equation}
Since the outgoing particle is described as an outgoing S-wave, the $WKB$ approximation can be applied to it. Particle emission rate and the imaginary part of action has the following relationship:
\begin{equation}
    \Gamma\propto e^{-2ImS},
\end{equation}
with $S$ is the action of the particle. The imaginary part of the action of the tunneling particle can be expressed as
\begin{equation}\label{q21}
\begin{aligned}
ImS&=Im\int_{r_{in}}^{r_{out}}P_{r}dr\\&=Im\int_{r_{in}}^{r_{out}}\int_{0}^{P_{r}}dP^{\prime}{}_{r}dr\\&=Im\int_{r_{in}}^{r_{out}}\int_{0}^{H}\frac{dH}{\bar{r}}dr,\end{aligned}
\end{equation}
where $P_r$ is the regular momentum corresponding to $r$, $r_{in}$ and $r_{out}$ correspond to the positions of particles at the moment before and after emission, respectively. We have used the Hamilton's equation
\begin{equation}
    \bar{r}=\frac{dH}{dP_r},
\end{equation}
Here, $dH$ characterizes the energy variation arising from radiation when the tunneling particle traverses the event horizon. In the conventional phase space framework, the black hole's internal energy $U$ is reinterpreted as the black hole mass $M$ . Given that the objective of this paper is to investigate the radiation behavior of particles in the extended phase space, where cosmological parameters are treated as dynamic variables, the black hole mass should be interpreted as the thermodynamic enthalpy $\mathcal{H}$ rather than the internal energy $U$ \cite{bib51}. Consequently, the energy change ought to be expressed as $d(M-\omega^{\prime})-Vd(P-p^{\prime})$. On this basis, Eq.(\ref{q21}) can be rewritten as follows.
\begin{equation}\label{q23}
    \begin{aligned}ImS=Im\int_{(M,P)}^{\left(M-\omega^{\prime},P-p^{\prime}\right)}\int_{r_{in}}^{r_{out}}\frac{dr}{1-\sqrt{\frac{2(M-\omega^{\prime})}{r}-\frac{8\pi(P-p^{\prime})}{3}r^{2}}}[d(M-\omega^{\prime})-Vd(P-p^{\prime})].\end{aligned}
\end{equation}
Obviously, there is a single pole at $r = r_H $ in above equation at the horizon, so, to calculate the above integration, we have to deform the contour around this pole. Set
\begin{equation}
    u=\frac{2(M-\omega^{\prime})}{r}-\frac{8\pi(P-p^{\prime})}{3}r^{2}.
\end{equation}
Simultaneous differentiation on both sides of the equation, we obtain
\begin{equation}
    \frac{r}{2}du=\mathrm{d}(M-\omega^{\prime})-\frac{4\pi r^3}{3}d(P-p^{\prime}).
\end{equation}
In the previous discussion, we have given the following.
\begin{equation}
    V=\frac{4\pi r^3}{3}.
\end{equation}
Which leads to
\begin{equation}
    \mathrm{d}(M-\omega^{\prime})-Vd(P-p^{\prime})=\frac{r}{2}du.
\end{equation}
And under this replacement, we get
\begin{equation}
    ImS=Im\int_{r_{in}}^{r_{out}}\int_0^u\frac{rdu}{2(1-\sqrt{u})}dr
\end{equation}
It can be seen that this integral is the existence of a pole at $u = 1$. Therefore, the above integral can be calculated using the contour integral, and we first integrate $u$, and integrating in the upper half-complex plane yields
\begin{equation}
    Im\int_0^\mathrm{u}\frac{rdu}{2(1-\sqrt{u})}=-i\pi r.
\end{equation}
In this way, the integration Eq.(\ref{q23}) can be easily written as
\begin{equation}
    \begin{aligned}ImS&=Im\int_{r_{in}}^{r_{out}}(-i\pi r)dr\\&=-\frac{\pi}{2}\left(r_{out}^{2}-r_{in}^{2}\right).\end{aligned}
\end{equation}
So the relationship betweend the tunneling rate can be obtain as 
\begin{equation}\label{q31}
    \Gamma\propto e^{-2ImS}=e^{\pi\left(r_{out}^{2}-r_{in}^{2}\right)}=e^{\Delta S_{BH}},
\end{equation}
where ${\Delta}S_{B H}=S_{B H}{(M-{\omega^\prime},P-p^\prime)-S_{B H}{(M,P)}}$ is the difference of the black hole entropy after and before the particle tunnels into the event horizon. The derived emission spectrum actually deviates from pure thermal in the extended phase space, which is consistent with that obtained in the normal phase space.
This result conforms to the unitary principle and supports the conservation of information.

%=========+=========+=========+=========+=========+=========+=========+=========
\section{Thermodynamic analysis of the particle tunneling radiation process}\label{sec:4}
In fact, in the context of both static spherically symmetric black holes and steady-state axially symmetric black holes, the expression of the first law of thermodynamics in reversible processes can be resorted to in verifying the conservation of information, regardless of whether the outgoing particles under discussion are massless, massive or charged particles. However, it is worth noting that this step has not been fully considered in de Sitter spacetime as well as in anti-de Sitter spacetime. 
Early literature has demonstrated a robust correlation between the emission spectrum of black holes and their first law of thermodynamics, yet this correlation lacks the inclusion of thermodynamic volume and pressure terms. Through the detailed analysis presented in the first section of this paper, it becomes evident that with the introduction of the extended phase space concept, the form of the first law of thermodynamics has undergone corresponding modifications and extensions, thereby incorporating the previously neglected thermodynamic volume and pressure terms. This refined formulation not only enriches the thermodynamic description of black holes but also paves the way for a more comprehensive exploration of the interplay between black hole radiation mechanisms and fundament, the
\begin{equation}\label{q32}
    d(M-\omega^{\prime})=T^{\prime} dS^{\prime}+V^{\prime}d(P-p^{\prime}),
\end{equation}
where $T$ is the Hawking temperature, and $T^{\prime}={{\kappa}^{\prime}}/{2\pi},\quad V^{\prime}=\frac{4}{3}\pi r_+^{\prime3}.$  Eq.(\ref{q32}) can be modified to
\begin{equation}\label{q33}
   dS^{\prime}=\frac{d(M-\omega^{\prime})-V^{\prime}d(P-p^{\prime})}{T^{\prime} }.
\end{equation}
Considering the asymptotic behavior of the outgoing particles near horizon, the geodesics equation of the outgoing particles reads
\begin{equation}\label{q34}
    \bar{r}\approx\kappa^{\prime}(r-r_h),
\end{equation}
here $\kappa^{\prime}$ is the surface gravity when the particle emitted from the horizon. Substituting Eq. (\ref{q33}) and Eq. (\ref{q34}) into Eq. (\ref{q23}), the imaginary part of the classical forbidden orbital action can be expressed as
\begin{equation}\label{q34}
    \begin{aligned}
    Im\mathcal{S}&\begin{aligned}\simeq Im\int_{r_{in}}^{r_{out}}\int_{(M,P)}^{(M-\omega,P-p^{\prime})}\frac{d(M-\omega^{\prime})-V^{\prime}d(P-p^{\prime})}{\kappa^{\prime}(r-r_+^{\prime})}dr\end{aligned}\\
    &=-\frac{1}{2}\int_{S_{BH}(M,P)}^{S_{BH}(M-\omega,P-p^{\prime})}dS^{\prime}=-\frac{1}{2}\Delta S_{BH}
    \end{aligned}.
\end{equation}
The probability of the outgoing particle crossing the event horizon is
\begin{equation}
    \Gamma\sim e^{-2\operatorname{Im}S}= e^{\Delta S_{BH}}.
\end{equation}
It is clear that with the help of the differential form of the first law of thermodynamics of black holes and the second law of thermodynamics, we are able to derive the probability that the outgoing particles crosses the event horizon of black hole and that this probability coincides with the result previously derived through equation Eq. (\ref{q31}). However, it must be made clear that this derivation only applies to reversible static processes. This is because, in a non-reversible process, the expression of the second law of thermodynamics will change and no longer retain its original form
\begin{equation}\label{q38}
    \begin{aligned}
    \mathrm{d}S>\frac{1}{T}(\mathrm{d}M-V\mathrm{d}P
    )\end{aligned}.
\end{equation}
This is due to the fact that the corresponding process has irreversible entropy generation, whereas substituting Eq. (\ref{q38}) for Eq. (\ref{q32}) will not give the result of Eq. (\ref{q34}). Anyway, this result is the same as the result obtained by Eq. (\ref{q31}), which on the other hand proves that even in the extended phase space, we add the consideration of the thermodynamic pressure, and the emission spectrum of the particles obtained actually deviates from the purely thermal spectrum, and the emission rate of the particles is related to the difference of the Bekenstein-Hawking entropy before and after the outgoing particles traverse the potential barriers, and at the same time consistent with an underlying unitary theory in the quantum mechanics, which supports the conclusion of the conservation of information.
%=========+=========+=========+=========+=========+=========+=========+=========
\section{Summary and discussion}\label{sec:5}
In summary, we try to investigate Hawking radiation in the extended phase space. The cosmological parameter $\Lambda$ was treated as a dynamic variable and the thermodynamic phase space was extended. By considering the cosmological constant as a thermodynamic pressure, although many interesting physical phenomena have been presented, Hawking radiation with thermodynamic pressure and volume remains unknown. 
On the basis of this, we employed the null geodesic method to carefully investigate Hawking radiation as tunneling of the AdS black hole, where the thermodynamic pressure and volume are taken into account. At first, we analyzed the geodesic motion of massless particles by employing the Lagrangian analysis method. With inclusion of the self-gravitation effect, we have used the geodesic equation to carefully investigate the tunneling radiation of massless and massive particles through the event horizon of the Schwarzschild Anti-de Sitter black hole in the extended phase space.  Finally, the results here show that the emission rate of emitted particles is proportional to a factor, which depending on the initial entropy and final entropy of black hole. 
This not only confirms the applicability of the tunneling process to scenarios involving thermodynamic pressure in the extended phase space, but also reveals its intrinsic connection with the first law of thermodynamics. Notably, from our thermodynamic analysis perspective, the relationship $\Gamma\sim e^{-2\Delta S}$ is a natural consequence of the first law of thermodynamics in the extended phase space (expressed in the form of $d M = TdS + P dV$), thereby reflecting a close interdependence governed by the principle of energy conservation.
This means that the exact emission spectrum also deviates from the pure thermal spectrum, even if we take the thermodynamic pressure into account. Therefore, it is true that our results are not only consistent with an underlying unitary theory, but also well coincide with that obtained in the normal phase space. In this paper, our discussion only focuses on the tunneling radiation of the Schwarzschild Anti-de Sitter black hole, it is of interest to discuss the Hawking radiation of other black holes in the extended phase space.

%=========+=========+=========+=========+=========+=========+=========+=========

\section*{Acknowledgments}
This work is supported by the National Natural Science Foundation of China (Grants Nos. 12375043), and National Natural Science Foundation of China (Grants Nos. 12575069 ).

%=========+=========+=========+=========+=========+=========+=========+=========

\end{document}